\documentclass[aps,prl,twocolumn,superscriptaddress]{revtex4-1}
\usepackage{xspace}
\usepackage[colorlinks]{hyperref}

\newcommand{\ket}[1]{\ensuremath{\left| #1 \right>}}

\newcommand{\braket}[1]{\ensuremath{\left< #1 \right>}}

\newcommand{\fixme}[1]{{\em\textcolor{blue}{#1}}}

\newcommand{\eg}{\emph{e.g.}\@\xspace }

\newcommand{\mb}{\mathbf}

\usepackage{float}
\usepackage{graphicx}
\usepackage{epstopdf}
\usepackage{setspace}
\usepackage{bbold}
\usepackage{comment}
\usepackage{color}
\usepackage{soul}
\usepackage[abs]{overpic}
\usepackage{amsmath}
\usepackage{lipsum}
\usepackage[abs]{overpic}
\usepackage{nicefrac}
\begin{document}

\author{Gadi Afek}
\affiliation{Wright Laboratory, Department of Physics, Yale University, New Haven, Connecticut 06520, USA}
\author{Daniel Carney}
\affiliation{Physics Division, Lawrence Berkeley National Laboratory, Berkeley, California 94720, USA}
\author{David C. Moore}
\affiliation{Wright Laboratory, Department of Physics, Yale University, New Haven, Connecticut 06520, USA}

\title{Coherent scattering of low mass dark matter from optically trapped sensors}

\begin{abstract}
We propose a search for low mass dark matter particles through momentum recoils caused by their scattering from trapped, nm-scale objects. Our projections show that even with a modest array of fg-mass sensors, parameter-space beyond the reach of existing experiments can be explored. The case of smaller, ag-mass sensors is also analyzed - where dark matter can coherently scatter from the entire sensor - enabling a large enhancement in the scattering cross-section relative to interactions with single nuclei. Large arrays of such sensors have the potential to explore new parameter space down to dark matter masses as low as 10~keV. If recoils from dark matter are detected by such sensors, their inherent directional sensitivity would allow an unambiguous identification of a dark matter signal. 
\end{abstract}

\maketitle
\paragraph*{Introduction.}
It is now evident from astrophysical observations that the majority of matter in the Universe consists of dark matter (DM), although its detection in the laboratory remains an outstanding challenge. Searches for weakly interacting massive particles (WIMPs) are among the most-developed techniques for terrestrial DM searches, employing multi-ton detectors~\cite{Schumann:2019eaa}. Despite the exquisite sensitivity of such detectors, no conclusive evidence for the existence of WIMPs has been reported to date.

DM particles could have evaded detection if they produce energy deposits below the threshold of existing detectors. Developing new techniques to achieve lower energy thresholds has thus recently become a major focus of the DM community~\cite{essig2020low}. Essentially all techniques proposed thus far seek to detect transfer of energy from the DM particle to a specific microscopic internal degree of freedom within a large detector~\cite{hochberg2016detecting,hochberg2017directional,knapen2018detection,budnik2018direct,Hochberg2018,Kurinsky2019,Hertel2019,Blanco2020,GriffinSiC2021,Hochberg2021}. A fundamentally different approach is to optically monitor the center-of-mass (COM) motion of a levitated macroscopic object, in order to detect small momentum transfers from the scattering of incident DM particles. Early proposals for WIMP and neutrino detectors considered this approach~\cite{shvartsman1982possibility,smith2003detection}, but required tracking the motion of a large array of individual, $\lesssim$~fg masses, then technologically infeasible. 


Following the pioneering work of Ashkin and Dziedzic~\cite{Ashkin:1971}, the modern development of levitated optomechanics has made substantial technical advances required to enable such ideas~\cite{Levitodynamics2021}. Levitated optomechanical sensors enabling sensitive searches for DM and other weakly coupled phenomena have been demonstrated~\cite{Monteiro2020DM,westphal2021measurement,Blakemore2021,AfekMCP2021} or proposed~\cite{smith2003detection,Riedel2013,bateman2015existence,Riedel2017,Carney2019_gravDM,timberlake2021probing,Cheng:2019vwy,Blanco2021models}. Extending such systems to large arrays of sensors --- a rapidly growing tool in the case of single atoms using optical traps~\cite{Lester2015,Endres2016,Norcia2018,Barredo2018,Covey2019} or ions using electromagnetic traps~\cite{zhang2017observation,Kumph_2011,Gilmore673}, and routine for fluid-levitated spheres~\cite{Roichman2007} --- could lead to substantial sensitivity improvements. 

\begin{figure} 
    \begin{centering}
    \includegraphics[width=0.45\textwidth]{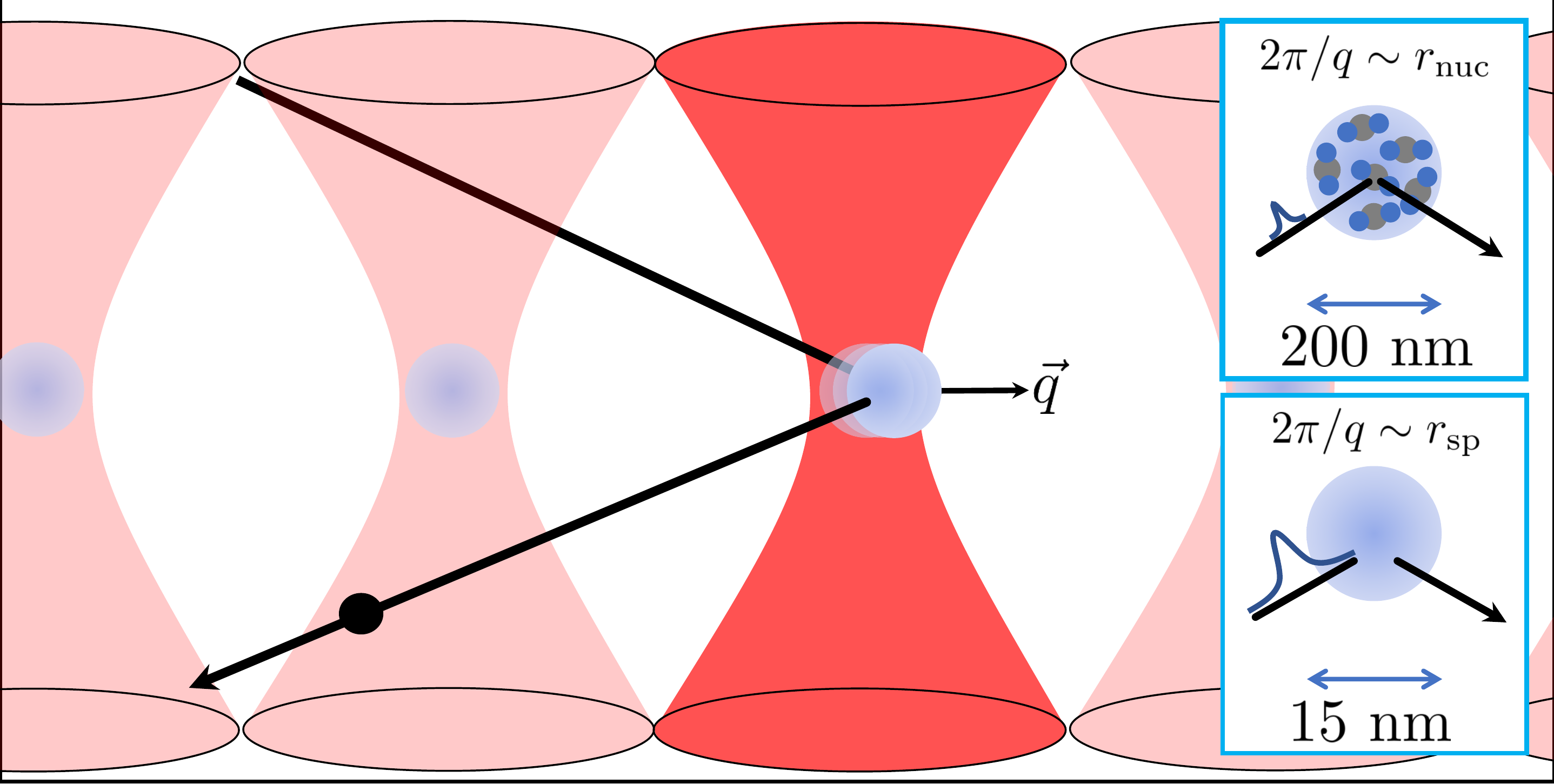}
    \caption{As a dark matter particle scatters from a levitated optomechanical sensor (possibly part of a large array), it transfers to it momentum $\vec{q}$. For ``large" sensors (upper inset) the interaction is coherent over a single nucleus. For ``small" enough sensors, such that the inverse transferred momentum $2\pi/q$ of the dark matter particle is comparable to the size of the sensor, the interaction is coherent over the entire sensor, leading to a large increase in scattering cross-section.}\label{Fig:Fig1}
    \end{centering}
\end{figure}

We propose the use of an array of nanoscale levitated sensors to search for DM with $\ll$~GeV mass. Monitoring the COM motional degrees of freedom of the particle allows measurement of the momentum transfer from the colliding DM even if it interacts only with a single nucleus in the sensor. The momentum transfer, which provides information about the daily modulation of the direction of colliding DM particles, can be measured in 3D (although even 1D is sufficient for the measurements proposed here~\cite{Monteiro2020DM}) at a noise level around the ``standard quantum limit" (SQL)~\cite{caves1981quantum,PhysRevB.70.245306}. The basic setup is depicted schematically in Fig.~\ref{Fig:Fig1}. 

At DM masses $\gtrsim10$~MeV~\footnote{DM particle masses, momentum transfers and detection thresholds in this paper are reported in natural units such that $\hbar=c=1$. Other parameters such as sphere masses are reported in SI units}, we consider $\sim$~fg-mass sensors, for which momentum sensitivity approaching the SQL has recently been demonstrated~\cite{Tebbenjohanns2020,Delic2020,Tebbenjohanns2021,magrini2021real}. This is sufficient to detect recoils below the energy threshold of existing DM direct detection experiments, while enough mass can be obtained to explore new parameter space with a modest-sized sensor array. For relic DM masses $\lesssim100$~keV, the momentum is low enough that they will exhibit coherent elastic scattering from a nm-scale, $\sim$~ag-mass object. This leads to a substantial increase in the cross section relative to that for a single nucleus or electron~\cite{Schumann:2019eaa}. While not yet experimentally demonstrated, reaching momentum sensitivity at (or beyond) the SQL in a sensor of this size would be sufficient to detect such coherent scatters, allowing new parameter space to be explored down to DM masses as low as 10~keV. This range of parameter space is, to date, entirely unexplored via direct detection.

\paragraph*{Scattering rate.}
We focus on spin-independent scattering of DM in the case of a heavy mediator with mass that is much larger than the momentum transfer $q$. The differential scattering rate per trapped sphere of mass $m_{\rm{sp}}$ and radius $r_{\rm{sp}}$ is then given by~\cite{Schumann:2019eaa}:
\begin{equation}
    \frac{dR}{dq} = \frac{\rho_\chi}{m_\chi}\frac{\sigma_{SI}}{2\mu^2}q\eta\left[v_{\rm{min}}(q)\right]S(q).
\label{eq:rate}
\end{equation}
Here $v_{\rm{min}}(q)=q/2\mu_{\chi T}$ is the minimum velocity for a given momentum transfer, $\mu_{\chi T}$ is the DM-target reduced mass, $\mu$ is the DM-nucleon reduced mass, $\sigma_{SI}$ is the single nucleon cross section, and $\eta(v)$ encodes the velocity distribution of the DM. We use standard assumptions about the virialized DM halo~\footnote{A truncated Maxwell-Boltzmann velocity distribution is assumed for the DM halo~\cite{Schumann:2019eaa}, with local circular velocity $v_0=220$~km/s, average Earth velocity $v_{\rm{E}}=245$~km/s, and galactic escape velocity $v_{\rm{esc}}=544$~km/s. The local DM density is assumed to be $\rho_\chi \approx0.3$~GeV/cm$^3$, and $\eta(v)$ is given in~ \cite{essig2016direct}.}. The function $S(q)$ contains details of the structure of the target that affect the scattering rate. In particular, for momentum transfers $q \lesssim 2\pi/r_{\rm sp}$, the scattering is quantum-mechanically coherent, and $S(q)$ grows \emph{quadratically} with the number of nuclei in the sphere, leading to substantial enhancement to sensitivity.

In more detail, for interactions with nucleons (assuming equal couplings to protons and neutrons), $S(q) = \sum_i A_i^2N_i F_{H}^2(q,A_i) + N_{\rm{n}}^2F_c^2(q)$. The first term gives the contribution from coherent scatters from nuclei in the target (which dominate at large $q$). If a target contains multiple species of nuclei then a sum is taken over each type $i$, where $N_i$ is the number of $i$-type nuclei in the sensor, $A_i$ is the respective mass number, and the Helm form factor is $F_{H}(q, A_i) = 3j_1(r_i q)\exp\left[-(sq)^2/2\right]/\left(r_i q\right)$~\footnote{For example, in a SiO$_2$ sphere, where each molecule has a mass of $m_{\rm{mol}}\approx 60$~amu, $N_{\rm{Si}} = m_{\rm{sp}}/m_{\rm{mol}}$ and $N_{\rm{O}} = 2\times m_{\rm{sp}}/m_{\rm{mol}}$.}. Here we assume $r_i=1.22~{\rm{fm}}\times A_i^{1/3}$ for the nuclear radius and $s=1$~fm for the skin depth~\cite{Schumann:2019eaa}. The second term in $S(q)$ dominates at sufficiently low $q$ that the substructure of the sphere cannot be resolved. For such low $q$, the form factor for coherent scattering from the entire sphere is given by $F_c(q) = 3j_1(r_{\rm{sp}} q)/\left(r_{\rm{sp}}q\right)$, where $j_1$ is the first order spherical Bessel function. $N_{\rm{n}}$ is the total number of nucleons in the sphere.

For an object trapped in a harmonic potential with trapping frequency $\omega$, a convenient benchmark for the detection threshold is the SQL for momentum impulses, $\sigma_{\rm{SQL}} = \sqrt{m_{\rm{sp}}\omega}$~\cite{caves1981quantum,PhysRevB.70.245306} (where $\hbar=1$~\cite{Note1}). For sub-wavelength objects, the minimal optically detectable impulse allowed by the Heisenberg uncertainty principle at the optimal readout laser power, compromising measurement backaction, laser shot-noise and detection efficiency is $\sigma_{\rm{SQL}}\left(2/5\eta_c\right)^{1/4}$, where $\eta_c$ is the total detection efficiency~\cite{Jain2016}. An in-depth discussion of the practical limits of $\eta_c$ can be found in~\cite{Tebbenjohanns2019Optimal}, indicating that the necessary efficiency can be achieved using a high numerical aperture imaging system in all 3 motional degrees of freedom. The SQL does not, however, represent a fundamental limit and it has been experimentally surpassed in recent years in a variety of systems~\cite{gross2010nonlinear,hosten2016measurement,rossi2018measurement,McCuller2020,backes2021quantum}.

\paragraph*{``Large" sensors -- nuclear coherence.}

\begin{figure*}
    \begin{centering}
    \includegraphics[width=\textwidth]{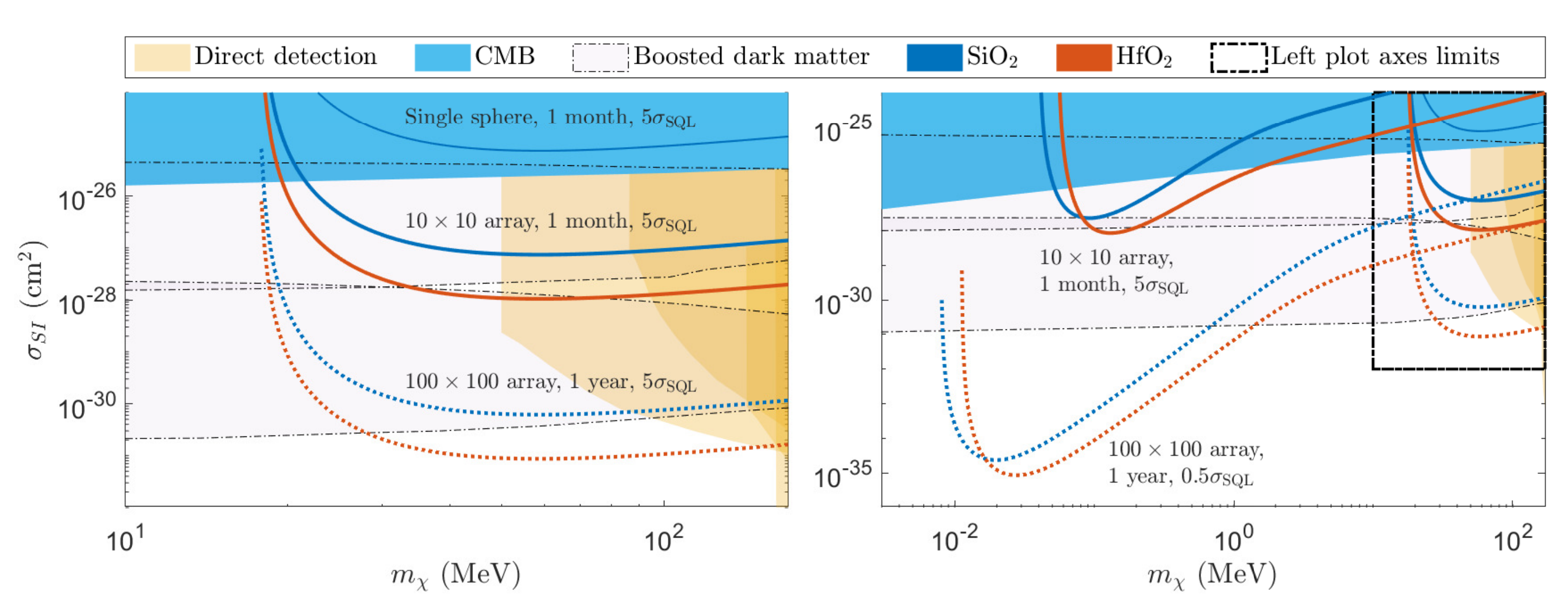}
    \caption{90\% CL sensitivity for recoil detection with nanosphere arrays, in terms of the single-nucleon cross section as a function of DM mass. (Left) ``Large", $200$~nm-diameter spheres, where DM interacts coherently with individual nuclei in the spheres. (Right) ``Small", 15~nm-diameter spheres, where the interaction is coherent over the entire sensor. Light yellow regions represent overlapping direct-detection limits~\cite{alkhatib2020light,angloher2017results,AbdelhameedCRESST2019,Collar2018}, with darker shades where multiple overlapping regions from different experiments exist, and blue regions indicate CMB-based constraints~\cite{Gluscevic2018}. The light gray regions indicate constraints on ``boosted DM", requiring upscattering of DM by cosmic rays~\cite{Bringmann2019,ProspectBoosted2021}. Constraints from meson experiments and cosmology also exist at masses $m_{\chi} \lesssim 100~{\rm MeV}$~\cite{Knapen2017,Krnjaic2020BBN}, as well as from galactic structure~\cite{NadlerDES2021}, but in general such constraints may depend on the specific DM model assumed.}\label{Fig:fig2}
    \end{centering}
\end{figure*}

For the case of a $\sim200$~nm-diameter sphere (upper inset of Fig.~\ref{Fig:Fig1}), where ground-state cooling of such (single) objects has recently been demonstrated~\cite{Tebbenjohanns2020,Delic2020}, Fig.~\ref{Fig:fig2} (left) shows an example of the projected sensitivity for the single-nucleon spin-independent scattering cross section, $\sigma_{SI}$, and DM mass, $m_\chi$ (thin blue solid line). The 90\% confidence level (CL) sensitivity is calculated for a single sensor and a month-long integration, assuming no observed events. Reaching this sensitivity requires the expected background rate in the assumed integration time to be $\ll 1$~event (see discussion below) above a threshold of $5\sigma_{\rm{SQL}}$ for $\omega = 2\pi\times 20$~kHz. The background-free sensitivity is compared for both SiO$_2$ (typically used in existing traps and readily commercially available) and HfO$_2$, which would have similar optical properties but a larger atomic number and mass density ($A = 178$ for Hf and a density of 9.86~g/cm$^3$ for HfO$_2$, compared to $A = 28$ for Si and 1.8~g/cm$^3$ for SiO$_2$). 

Such an initial search using a single sphere with a month-long integration would already be projected to reach lower DM masses, and similar cross-sections for $m_\chi \lesssim 100$~MeV than existing direct detection constraints at these masses (\eg SuperCDMS~\cite{alkhatib2020light} and CRESST~\cite{angloher2017results,AbdelhameedCRESST2019} as well as~\cite{Collar2018}) and cosmological constraints~\cite{Gluscevic2018}. Lower cross sections can be reached by considering an array of such spheres and longer integration times. Examples of sensitivity curves for a $10\times10$ array with a month-long integration and a $100\times100$ array with a year-long integration are shown in thick solid and dashed lines. 

\paragraph*{``Small" sensors -- full coherence}

If the sensor is chosen such that its radius is comparable to the inverse momentum transfer, $2\pi/q$, the interaction becomes coherent over the entire sphere. Uniquely, for such few-nm objects the momentum detection sensitivity can reach their inverse size, while maintaining sufficient sensor mass to reach relevant cross sections. For SiO$_2$ nanospheres with SQL detection sensitivity and a trapping frequency of $2\pi\times 1$~kHz, the optimal sensor size occurs at a diameter of 15~nm, giving $\sim10^6$ nucleons in a sphere and $\sim 12$ orders of magnitude enhancement in the scattering cross section compared to a single nucleon. Such objects are commercially available and may be trapped optically or  electromagnetically~\cite{zhang2017observation,Kumph_2011,Gilmore673}. 

Fig.~\ref{Fig:fig2} (right) shows the projected single-nucleon cross section sensitivity of this fully-coherent case. The smaller spheres enable a reduced momentum threshold (again assuming 5$\sigma_{\rm{SQL}}$), lowering the detectable DM mass to $\lesssim100$~keV for the example discussed here. Trapping a large array, assuming the same sensitivity, or alternatively beyond-SQL detection may allow sensitivity to cross sections approaching the picobarn level. The dotted lines in Fig.~\ref{Fig:fig2} (right) indicate the sensitivity possible with an array reaching a detection threshold of 0.5$\sigma_{\rm{SQL}}$.

Coherent scatters from the sphere can also be detected if DM primarily interacts with electrons rather than nucleons. Since these techniques are sensitive to energy transfers below the threshold for ionization~\cite{Essig2012,Barak2020,DAMIC:2020cut,EDELWEISS2020,Amaral2020}, DM masses down to 10~keV can be probed. However, the proposed techniques using electrically neutral particles are sensitive only if DM does not couple to total electric charge but rather to electron number~(\eg, \cite{2009PhRvD..79h3528F}), since the total charge of a neutralized sphere is zero, and coherent scattering cannot probe the substructure of the charge distribution.

\paragraph*{Backgrounds.}

The analysis presented above assumes that backgrounds can be sufficiently identified and rejected to ensure $\ll 1$~ expected background event in the required integration times. For the backgrounds identified in the following section, this appears to be plausible, although verifying it will require further investigation. As with any new technique, unexpected backgrounds are possible and would need to be studied in realistic implementations. The nature of the backgrounds identified here may allow operation in surface laboratories (or possibly even space). In fact, recent proposals for space missions employing trapped nanoparticles for tests of quantum mechanics may also enable such searches~\cite{Riedel2013,Riedel2017,MAQRO2021}.

A significant background may arise from the residual gas present in ultra-high vacuum (UHV) environments, whose collisions with the sphere can transfer momentum similar to the expected DM signal. The rate of such collisions strongly depends on the ambient pressure, while the momentum transfer per collision depends on the temperatures of the gas and the sphere surface. Three different vacuum pressures are analyzed: $10^{-9}$~mbar (achievable with mechanical pumping), $10^{-12}$~mbar (ion pumps) and $10^{-15}$~mbar (record pressures achieved in cryogenic systems~\cite{Gabrielse1990}). 

\begin{figure}
    \begin{centering}
    \includegraphics[width=0.5\textwidth]{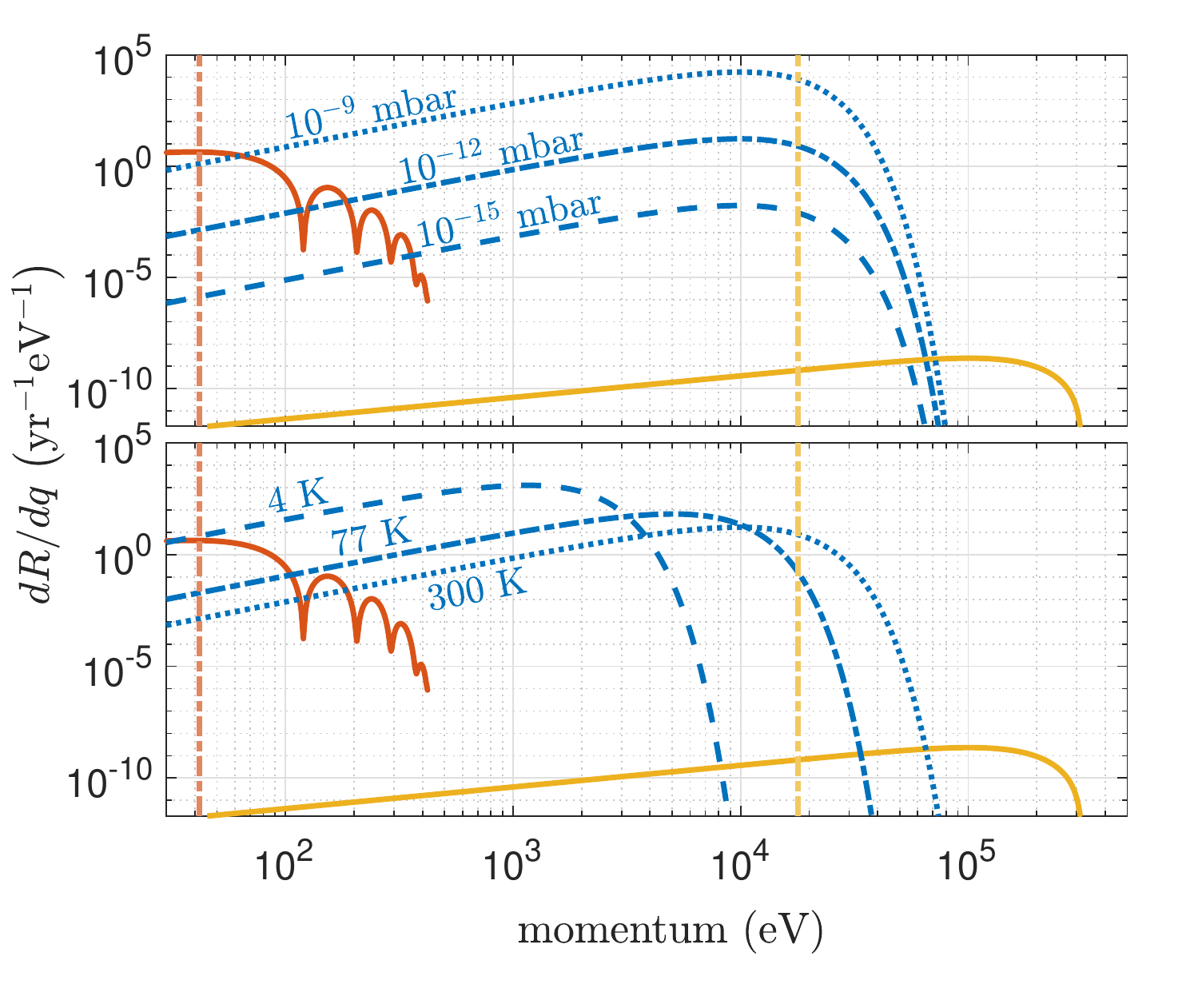}
    \caption{Expected thermal backgrounds for different vacuum pressure and ambient temperatures. The differential scattering rate (Eq.~\ref{eq:rate}) for the ``large" (yellow, high momentum) and ``small" (red, low momentum) sensors is compared to the simulated recoil spectrum from background gas. The top panel indicates the expected spectra at a sphere and ambient temperature of 300~K for different vacuum pressures. The bottom panel varies the temperature at  $10^{-12}$~mbar pressure. The dash-dotted vertical lines correspond to the respective momentum detection thresholds.}\label{Fig:fig3}
    \end{centering}
\end{figure}

Even at the highest pressures considered, the mean time between collisions with residual gas particles is sufficiently long compared to $\omega^{-1}$ that these can be treated as independent, isolated events. The collisional recoil spectrum is calculated using a Monte Carlo simulation under the assumptions that the dominant residual gas is H$_2$, which is diffusely reflected from the particle~\cite{COMSA1985145}. In this model, incident gas particle velocities are drawn from a Maxwell distribution at ambient temperature. Upon collision with the sphere, gas particles are diffusely emitted with a velocity distribution given by the sphere temperature and a $\cos{\theta}$ angular distribution~\cite{COMSA1985145}. For each simulated collision the total momentum transfer is calculated and the resulting spectrum is fit with a Maxwell profile. The resultant spectra are shown in blue dotted, dashed and dash-dotted lines in Fig.~\ref{Fig:fig3}. 

For the ``large" sphere case, the expected DM signal (solid line) and threshold (dash-dotted vertical line) are plotted in yellow, assuming a 200~nm-diameter sphere with DM mass of 60~MeV and $\sigma_{SI}=6\times10^{-31}~{\rm cm}^2$, corresponding to the lowest point of the blue dotted line in Fig~\ref{Fig:fig2} (left). At room temperature the expected DM spectrum extends to higher momentum than the thermal gas distribution, and cooling the gas and the sphere surface is shown to further reduce backgrounds. In contrast, for the ``small" sphere case [red, assuming a 15~nm-diameter sphere with $m_\chi=80$~keV and $\sigma_{SI}=2\times10^{-28}$~cm$^2$, corresponding to the lowest point of the blue solid line in Fig~\ref{Fig:fig2} (right)], the expected DM spectrum lies on the extreme low-momentum side of the thermal distribution. Cooling therefore does not help [Fig.~\ref{Fig:fig3} (bottom)], but reduction of the pressure does [Fig.~\ref{Fig:fig3} (top)]. The DM signal would then exceed the expected thermal gas collision rate near threshold for a pressure of $10^{-9}$~mbar for the 10$\times$10~sensor arrays, while lower pressure would be required to reach lower $\sigma_{SI}$ with larger arrays. These estimates rely on extrapolation of the thermal distribution to its extreme tails, and further investigation is required to test whether collisions follow the distribution assumed here in these regimes. Outgassing of molecules from the sphere itself may also contribute to the recoil background and will need to be investigated.

The dominant backgrounds in most existing DM searches arise from particle interactions of radiogenic or cosmogenic origin. Typically such interactions deposit much higher momenta than those of interest here. However, lower-energy secondary particles produced in conjunction with higher energy particle interactions in materials surrounding the sensors could introduce backgrounds. While any particle interaction changing the net charge of the sphere by even a single $e$ could be easily vetoed~\cite{Ashkin1980,Moore2014,Frimmer2017,Quidant:2019,Monteiro2020uK,AfekMCP2021}, a thermal neutron or low energy x-ray could induce momentum transfers in the eV - keV range of interest, without altering the charge state of the sphere. Such particles can also coherently scatter from the spheres, producing a signature identical to the DM signal of interest with corresponding enhancement in rate. However, the expected terrestrial flux of DM at low masses (\eg $\approx 10^{12}$~cm$^{-2}$\,sec$^{-1}$ for $m_\chi = 100$~keV) is much higher than the expected rate of such backgrounds, even after accounting for the significantly higher cross-sections. These secondary particles can be additionally vetoed by positioning conventional particle detectors around the trap to detect the higher energy primary particles or thermal neutron captures that would occur in coincidence with lower energy secondaries.

Fluctuations in blackbody radiation emitted by the sphere can also cause recoils. The expected momentum noise~\cite{Chang2010} is $\approx 3$~eV/$\sqrt{\rm{Hz}}$ for a (sub-wavelength) 15~nm sphere at 300~K, which is sub-dominant at the $\omega^{-1} \approx 1$~msec integration times assumed here.

Technical sources of noise such as vibrations can be significant. In an initial search for recoils of ng-mass spheres from scattering of heavy DM particles~\cite{Monteiro2020DM}, sources of such vibrational noise were found to be dominant, but could be effectively vetoed with a commercial accelerometer placed outside the vacuum chamber. An array of sensors can provide substantial additional reduction by vetoing impulses correlated among multiple sensors. 

The described system is sensitive to not only the amplitude of the momentum transfer but also to its direction by monitoring the recoil of the entire sphere in 3D (a projection onto 1D would also give directional sensitivity albeit with lower efficiency due to reduced angular acceptance). The direction of the recoil is predicted to modulate daily due to the change in the incident direction of the dark matter~\cite{MAYET20161}, and momentum conservation guarantees the sphere recoil matches the momentum transfer from the DM, even in the case that the DM initially produces only a nuclear recoil within the sphere. 

While unanticipated backgrounds may arise since this is an entirely new technology, this inherent directional sensitivity allows an unambiguous separation of a signal from any of the backgrounds described above.

\paragraph*{Trapping and detection.}
Optically trapping increasingly small sub-wavelength dielectric objects requires relatively high laser power to overcome thermal forces. A 100~nm SiO$_2$ sphere trapped with a $5~\mu$m waist, 1064~nm laser, for example, would need $\sim 100$~mW/trap. While the required optical power is achievable even for large arrays of such objects, for 15~nm spheres, the lower trap depth may make optical trapping impractical and necessitate the use of RF electromagnetic (``Paul") traps or Penning traps to confine the particles. Such traps are a scalable platform for trapping large numbers of objects~\cite{zhang2017observation,Kumph_2011,Gilmore673}, and ongoing work to extend to even larger arrays is driven by quantum computing efforts. Optical detection at the SQL requires detection efficiencies approaching unity. For sub-wavelength scatterers, practical upper limits in a high numerical aperture system are $\eta_c \lesssim 0.6$~\cite{Tebbenjohanns2019Optimal}, although further work is required to determine if such efficiencies can be reached in trap designs supporting a large array of particles.

\paragraph*{Summary.} We have suggested a new class of searches for low-mass, particle dark matter using levitated, nanoscale mechanical devices operated around the standard quantum limit for impulse sensing. These devices are capable of directional searches for DM masses in the keV--GeV regime, which has few current direct detection constraints, and are complementary to other proposals. Remarkably, due to coherent interaction at low momentum detection thresholds, sensitivities for probing new parameter space for DM can be achieved with sensor masses as small as ag--fg. Beyond the DM context, such sensors will be sensitive enough to count individual collisions of latent gas in ultra-high vacuum environments, possibly enabling an absolute pressure standard at ultralow pressures, a target of increasing importance in diverse fields of physics and metrology~\cite{scherschligt2017development}. In the search for low-mass DM, these nanoscale devices provide a plausible scheme to leverage quantum-coherent scattering of dark matter from a macroscopic target.

\begin{acknowledgments}
\paragraph*{Acknowledgments.} The authors would like to thank Simon Knapen, Gordan Krnjaic, and Jess Riedel for discussions. DC is supported by the US Department of Energy under contract DE-AC02-05CH11231 and the DOE QuantISED program. GA and DCM are supported, in part, by the Heising-Simons foundation and NSF Grant PHY-1653232.

\end{acknowledgments}

\bibliography{LowMassDM}

\begin{appendix}

\section{Appendix - review of coherent scattering}

Here we briefly review the basics of coherent enhancements to scattering on a many-body object. The treatment essentially follows~\cite{Riedel2013}; see also~\cite{LEWIN199687} for a more traditional exposition.

In standard potential scattering, an incoming plane wave incident on a radially symmetric potential evolves to the outgoing state
\begin{align}
\ket{\mb{p}} \to S \ket{\mb{p}} = \ket{\mb{p}} + \int d^3\mb{p}' \delta(E_{\mb{p}} - E_{\mb{p}'}) f(\mb{p},\mb{p}') \ket{\mb{p}'}.
\end{align}
Here $f(\mb{p},\mb{p}')$ is the scattering amplitude and $S$ is the usual $S$-matrix. If the potential had been displaced away from the origin to some location $\mb{x} \neq 0$, we would replace $S \to U S U^{\dagger}$ where $U = \exp(-i \mb{x} \cdot \mb{p})$ is the translation operator. Using this instead, the plane wave scatters to
\begin{align}
\begin{split}
\ket{\mb{p}} & \to U S U^{\dagger} \ket{\mb{p}} \\
& = \ket{\mb{p}} + \int d^3\mb{p}' \delta(E_{\mb{p}} - E_{\mb{p}'}) e^{-i \mb{x} \cdot \mb{q}} f(\mb{p},\mb{p}') \ket{\mb{p}'}
\end{split}
\end{align}
where the momentum transfer is $\mb{q} = \mb{p}' - \mb{p}$.

Now suppose that the incoming particle scatters off a sum of $N$ identical potentials, each located at a position $\mb{x}_i$. To lowest order in perturbation theory, the amplitude adds coherently, and the scattering process produces
\begin{align}
\ket{\mb{p}} \to \ket{\mb{p}} + \sum_{i=1}^{N} \int d^3\mb{p}' \delta(E_{\mb{p}} - E_{\mb{p}'}) e^{-i \mb{x}_i \cdot \mb{q}} f(\mb{p},\mb{p}') \ket{\mb{p}'}.
\end{align}
Here $f$ still represents the single-site scattering amplitude, computed to first Born order. (At higher orders in the Born approximation, the cross-terms between potentials will spoil the linear expression above.) The probability of a given outgoing momentum state $\ket{\mb{p}'}$ is therefore
\begin{align}
\label{coherent-basic}
P(\mb{p}') \propto \left| \braket{ \mb{p}' | \psi} \right|^2 \propto \sum_{ij} \braket{e^{-i \Delta \mb{x}_{ij} \cdot \mb{q}}}_{\rm target} | f(\mb{p},\mb{p}') |^2
\end{align}
away from the forward direction $\mb{q}=0$. Here $\Delta \mb{x}_{ij} = \mb{x}_i - \mb{x}_j$. The expectation value is taken over the internal states of the $N$-body target, whose state must be traced to get the inclusive probability for the scattered wave.  If the phases are all essentially random, the sum is incoherent and only the $N$ diagonal $i=j$ elements survive. Coherent scattering, on the other hand, is the situation that the momentum transfer is small compared to the inverse spacing between the potentials, i.e. $\Delta \mb{x}_{ij} \cdot \mb{q} \ll 1$ for all $i,j$, so that every phase in the sum is nearly the same, leading to a coherent enhancement. 

In more detail, we assume that the target body is in a thermal state with the particles distributed homogeneously within a sphere of radius $R$, so that
\begin{align}
\begin{split}
\label{phases}
\braket{e^{-i \Delta \mb{x}_{ij} \cdot \mb{q}}}_{\rm target} & = \frac{1}{(4\pi R^3/3)^2} \int d^3\mb{x}_i d^3\mb{x}_j e^{-i \Delta \mb{x}_{ij} \cdot \mb{q}} \\
& = F^2(qR),
\end{split}
\end{align}
with
\begin{align}
F(x) = \frac{3}{x^3} (\sin x - x \cos x).
\end{align}
Finally, we need to be a bit careful to make sure that the incoherent limit $qR \gg 1$ behaves correctly. In this limit, only the diagonal $i=j$ terms in \eqref{coherent-basic} survive. Modeling the target as a substance with $N_A$ sites each containing $A$ nucleons, we can write
\begin{align}
\begin{split}
\sum_{i,j=1,\ldots,N_A} \braket{e^{-i \Delta \mb{x}_{ij} \cdot \mb{q}}} & = A^2 \left[ \sum_{i} \braket{1} + \sum_{i \neq j} \braket{e^{-i \Delta \mb{x}_{ij} \cdot \mb{q}}} \right] \\
& = A^2 \left[ N_A + (N_A^2-N_A) F^2(qR) \right],
\end{split}
\end{align}
making use of our result \eqref{phases}. For small $x = qR \ll 1$, the form factor $F^2(x) \to 1$, while at large $x$, we have $F^2(x) \sim x^{-4}$. Thus at large momentum transfer the scattering probability scales like $N_A$, while at low momentum transfer we get a scaling like $N_A^2$. This also explains the prefactor $A^2$: we assume coherence over at least the single-site set of neutrons. All told, this means that we can write the total differential cross-section
\begin{align}
\frac{d\sigma}{dq} =  A^2 \left[ N_A + (N_A^2-N_A) F^2(qR) \right] \frac{d\sigma_{SI}}{dq},
\end{align}
where $d\sigma_{SI}$ represents the differential scattering cross-section on a single neutron.

In our problem, we have a solid body comprised of $N$ sites (say, SiO$_2$ molecules), each of which has $A$ neutrons and linear size $a \sim {\rm few}\times a_0$ with $a_0$ the Bohr radius. At the farthest reaches of our suggested approach, we are looking to detect impulses $q$ from an incoming DM particle of mass around $m_\chi \sim 10~{\rm keV}$, virialized to the galaxy $v \sim 10^{-3}$. This corresponds to a coherence length scale $q^{-1} \sim ({10^{-2}~{\rm keV}})^{-1} \sim 20~{\rm nm}$. These impulses should thus generate completely coherent scattering across the target. At DM masses around $1~{\rm MeV}$, the coherence length is down to $0.2~{\rm nm} \sim 4 a_0$, and so coherence applies only over one to a few nuclei. This effect leads to the bend in our sensitivity curves [Fig.~\ref{Fig:fig2} (right)].

\section{Calculation of thermal backgrounds}

To obtain the distribution of momenta transferred to the sphere by collisions with ambient H$_2$ gas molecules of mass $M_G$ within the vacuum chamber we assume that the gas has a Maxwell-Boltzmann velocity distribution with some temperature $T_G$. An incoming gas molecule bearing momentum $\mb{q}_{in}$ then collides with the sphere at some angle $\theta_{in}$ drawn from a $\cos(\theta)$ distribution. Assuming an interaction model of diffuse reflection~\cite{COMSA1985145}, the gas molecule then is adsorbed onto the sphere and thermalizes with it before being re-emitted at some $\cos(\theta)$-distributed polar angle $\theta_{out}$ and uniformly-distributed azimuthal angle $\phi_{out}$. From thermal equilibrium considerations, the magnitude of the outgoing momentum of the gas molecule $q_{out}$ is also Maxwell-Boltzmann distributed, but with the temperature of the sphere $T_S$ which is, in general, higher than $T_G$~\cite{Monteiro2017} \fixme{[cite novotny]}.

The calculation of the transferred momentum is performed using a Monte-Carlo simulation, drawing $3\times10^5~\mb{q}_{in}$ momentum components and respective $\theta_{in}$, $\theta_{out}$ and $\phi_{out}$ angles. For each of the $10^5$ collision events a transferred momentum $\mb{q}_{out} - \mb{q}_{in}$ is calculated and the resultant distribution $P(q)$ fitted to a Maxwell-Boltzmann distribution and rescaled according to $dR/dq = n\sigma vP(q)$. Here the number density $n$ is calculated using the ideal-gas law given the gas temperature $T_G$ and the desired pressure, $\sigma = \pi R_{\rm{Sp}}^2$ is the geometric cross section of the sphere and $v=\sqrt{8k_BT_G/(\pi M_G)}$. 

\end{appendix}

\end{document}